\documentclass[usenatbib]{mn2e}
\input epsf
\pagerange{}\volume{}\pubyear{}\date{}
\def\wind{w}
\begin{document}

\title[ Maximum feedback in dwarf galaxies ]
      { Maximum Feedback and Dark Matter Profiles of Dwarf Galaxies }

\author[ Gnedin \& Zhao ]
       { Oleg Y. Gnedin and HongSheng Zhao \\
         Institute of Astronomy, Cambridge CB3 0HA \\
         {\tt ognedin@ast.cam.ac.uk,  hsz@ast.cam.ac.uk}
       }
\maketitle

\begin{abstract}
The observed rotation curves of dark matter-dominated dwarf galaxies
indicate low density cores, contrary to the predictions of CDM models.
A possible solution of this problem involves stellar feedback.  A
strong baryonic wind driven by vigorous star formation can remove a
large fraction of the gas, causing the dark matter to expand.  Using
both numerical and analytical techniques, we explore the maximum
effect of the feedback with an instantaneous removal of the gaseous
disk.  The energy input depends on the compactness of the disk, hence
the specific angular momentum of the disk.  For the plausible
cosmological parameters and a wide range of the disk angular momenta,
the feedback is insufficient to destroy the central halo cusp, while
the inner density is lowered only by a modest factor of 2 to 6.  Any
realistic modeling of the feedback would have even lesser impact on
dark matter.  We find that no star formation effect can resolve the
problems of CDM cusps.
\end{abstract}


\section{Introduction}
  \label{sec:intro}

Dwarf spheroidal galaxies are excellent systems for testing current
theories of dark matter.  The large mass-to-light ratios indicate that
even their centers are dominated by dark matter \citep{M:98}.  The
velocity dispersion profile of stars and the rotation curve of gas
provide clean measures of the dynamical mass at all radii.  Recent
observations of the gas-rich dwarfs show a nearly solid-body rotation
curve, which implies a finite core inside a few hundred pc
(\citealt{vdB:00}; \citealt{deB:01}).  This is in conflict with the
predictions of the cold dark matter (CDM) models.  Numerical
simulations invariably produce a diverging, power-law density ``cusp''
$\rho \propto r^{-\gamma}$ with $\gamma = 1$ (\citealt{NFW:97}) or
$\gamma = 1.5$ (\citealt{Moore:99b}) within $\sim 500 \, (M_{\rm
vir}/10^9M_\odot)^{1/2}$ pc (\citealt{Bullock:01a}).

The additional disagreements of the CDM models with observations
include (1) the prediction of too many dwarf satellites within large
halos (\citealt{Ketal:99}; \citealt{Moore:99a}); (2) the triaxiality
of halos in conflict with the almost spherical cores of some clusters
of galaxies, as inferred from gravitational lensing \citep{Tetal:98};
(3) the efficient transport of the baryon angular momentum, leading to
too small disks of spiral galaxies \citep{NS:00}.  Of these, the
central cusps seem to be the most severe discrepancy.  The CDM models
have been very successful in matching observations on scales larger
than a Mpc, but the smaller-scale problems have lead to a recent
search for alternatives to CDM.  The variants include warm dark matter
(e.g., \citealt{BOT:01}) and self-interacting dark matter
(\citealt{SS:00}).

We examine a possible astrophysical solution to the cusp problem, the
effect of star formation feedback.  The energy released in supernova
explosions may heat and ionize the surrounding gas.  In the event of
an extremely powerful burst of star formation, the heated gas may
leave the dwarf galaxy in a form of fast wind (e.g., \citealt{DS:86}).
In an idealization of this problem, a significant fraction of the gas
may be removed from the dwarf halo on a timescale shorter than the
dynamical time.  We consider the reaction of the dark matter
distribution to the sudden loss of baryonic mass in the center.
Supermassive black holes can also have interesting effects, somewhat
different from the gas disks, and these will be discussed elsewhere.

\section{Cosmological context}
  \label{sec:para}

We explore the maximum effect of stellar feedback on the dark matter
profile, assuming (1) all cool gas within the dark halo can be
removed; (2) the amount of gas locked in stars is negligible; (3) the
removal of gas is instantaneous.

The initial conditions for the dark matter halo and the baryonic disk
are set using the cosmological concordance model with $\Omega_m =
0.35$, $\Omega_\Lambda = 0.65$, $\Omega_b \, h^2 = 0.02$
(\citealt{BOPS:99}), where $h = 0.65$ is the Hubble constant in units
of 100 km s$^{-1}$ Mpc$^{-1}$.  This gives the average baryon fraction
$f_b \equiv \Omega_b/\Omega_m \approx 0.13$.  Some of the baryons
associated with the halo of mass $M_{\rm vir}$ will never cool and
collapse towards the center, some may be heated and ionized early on
and escape the shallow potential well before the starburst.  But since
we are interested in the maximum effect, we assume that all the
associated baryons settle into a disk with the mass $M_d = f_b\,
M_{\rm vir}$.

The size of the rotationally-supported disk, $r_d$, is determined by
the angular momentum of the baryons (e.g., \citealt{MMW:98}).
Cosmological simulations show that dark matter halos acquire a
log-normal distribution of the dimensionless angular momentum,
$\lambda \equiv J |E|^{1/2}/G M_{\rm vir}^{5/2}$, with the median
$\bar{\lambda} = 0.05$ and dispersion $\sigma_\lambda = 0.5$
(\citealt{BE:87}; \citealt{EFWD:88}).  If the gas had the same angular
momentum as the halo, the disk size would scale with $\lambda$ ($r_d
\sim \lambda^2 \, r_{\rm vir}$ if disk is fully self-gravitating, and
$r_d \sim \lambda \, r_{\rm vir}$ in the no self-gravity case).

Initially, in isolated halos the gas is supposed to acquire the same
angular momentum per unit mass as the dark matter.  As a result of
hierarchical merging, the distribution of dark matter can become
highly non-homogeneous.  Also, the cooling and fragmentation of gas
into dense clouds can transfer orbital angular momentum to the dark
matter via dynamical friction and direct torques \citep{NB:91}.  In
the end, the gaseous disk may have a significantly lower value of the
parameter $\lambda_b \ll \lambda$.  We mimic the effect of angular
momentum redistribution by considering three model disks with
different $\lambda_b$.

\begin{figure}
\epsfxsize=8.4cm
\epsfbox{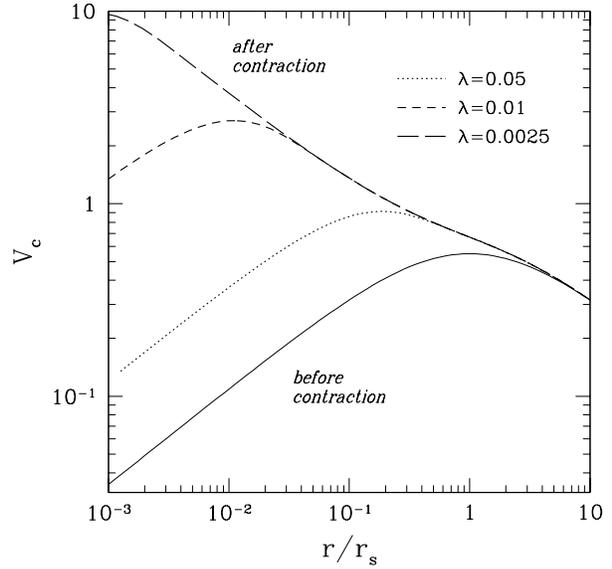}
\caption{Rotation curve before and after the formation of baryonic disk
  and adiabatic contraction of dark matter, for three values of the
  angular momentum parameter, $\lambda = 0.05, 0.01, 0.0025$.}
\label{fig:vel_init}
\end{figure}

The dark matter distribution self-consistently adjusts to the
formation of the disk in the middle of the halo.  This process can be
described as adiabatic contraction (\citealt{BFFP:86};
\citealt{Fetal:93}; \citealt{DSS:97}), which we discuss at length in
\S\ref{sec:adiabat}.  Such condensation of both baryons and dark
matter has a significant effect on the central density.  Figure
\ref{fig:vel_init} demonstrates how adiabatic contraction increases
the rotation curve in the inner galaxy over that predicted by
cosmological simulations and intensifies the discrepancy with
observations.

After the gas blowout, the final distribution of dark matter will
depend on the initial disk size in the following way.  If the disk was
initially relatively extended ($r_d \ga r_s$) the re-expansion phase
is driven by the energy input, which is proportional to the binding
energy of the disk, $\propto M_d/r_d$.  If the disk was very compact
($r_d \ll r_s$) only the total removed mass $M_d$ affects the results.
The new equilibrium profile after the removal of the entire baryonic
disk must be more extended than that prior to adiabatic contraction,
because any rapid expansion increases the actions of dark matter
orbits.  The faster the expansion, the stronger is the effect.

The energetics of gaseous winds have been studied extensively in the
past (\citealt{DS:86}; \citealt{vdB:00} and references therein).  In
general, dwarf galaxies have shallow potential wells and it is easy to
create a strong wind after a few supernova explosions.  We shall simply
assume that the SNe ejecta energy is sufficient and is efficiently
deposited into the gas, to facilitate a rapid expulsion of nearly all
the initial baryons with very little locked into stars and stellar
remnants.

\section{Analytical models}

\subsection{Two models of adiabatic contraction}
  \label{sec:adiabat}

We consider the problem in a spherically-symmetric approximation.  For
the initial virialised halo model we use a \citet{H:90} profile, which
is a special case of a family of analytical double-power-law models
(\citealt{Zhao:96}) with the mass $M_{dm}(r_i)$ enclosed within radius
$r_i$ given by
\begin{equation}
{M_{dm}(r_i) \over (1-f_b) M_{\rm vir}}
  = \left({r_i \over r_s+r_i}\right)^{3-\gamma},~~\gamma=1,
\end{equation}
where $r_s$ is the scale radius, and $\gamma$ is the slope of the
cusp.  We adopt a concentration parameter $r_{\rm vir}/r_s = 10$,
appropriate for the dwarf halos at high redshift, but the results at
small radii are insensitive to the actual value of this parameter.
The profile has the same inner slope ($\gamma=1$) as the NFW model
\citep{NFW:97} and deviates from it only in the outer parts which are
not interesting for the present problem.  The Hernquist model is more
convenient both for analytical modeling and for generating the initial
condition of an isolated fixed-mass halo.

During adiabatic contraction, circular orbits conserve the absolute
value of the angular momentum, $G M(r) r$, and the enclosed dark
matter mass, $M_{dm}(r)$.  An orbit can be labeled interchangeably by
its initial radius $r_i$ or by the specific angular momentum $j$,
where $j$ and $r_i$ are related by
\begin{equation}\label{j}
j^2 = r_i G M_{dm}(r_i) (1-f_b)^{-1}.
\end{equation}
Here we use the constant factor $(1-f_b)^{-1}$ to include the mass of
baryons assumed to have the same initial distribution.  The orbit with
an angular momentum $j$ has a post-contraction radius $r_j$ given by
\begin{equation}\label{rj}
r_j = {j^2 \over G M_{dm}(r_i) + G M_b(r_j)},
\end{equation}
where $M_b(r_j)$ is the mass of baryons enclosed within the contracted
orbit, which we need to specify separately.

We will use two independent models to evaluate adiabatic contraction.

1)
Using the observations of disk galaxies as a guide, it is common to
assume that the disk would have an exponential surface density
profile.  The enclosed disk mass is
\begin{equation}\label{exp}
M_{b,\rm exp}(r_j) = f_b M_{\rm vir} \left[1-(1+y) \exp(-y)\right],
  \qquad y \equiv {r_j \over r_d},
\end{equation}
where the scale length $r_d$ is computed numerically by matching the
total angular momentum (e.g., \citealt{MMW:98}, their eq. 28).  In this
approach, $r_j$ is an implicit function of $r_i$ or $j$, and the
equation must be solved iteratively.

\begin{figure}
\epsfxsize=8.4cm
\epsfbox{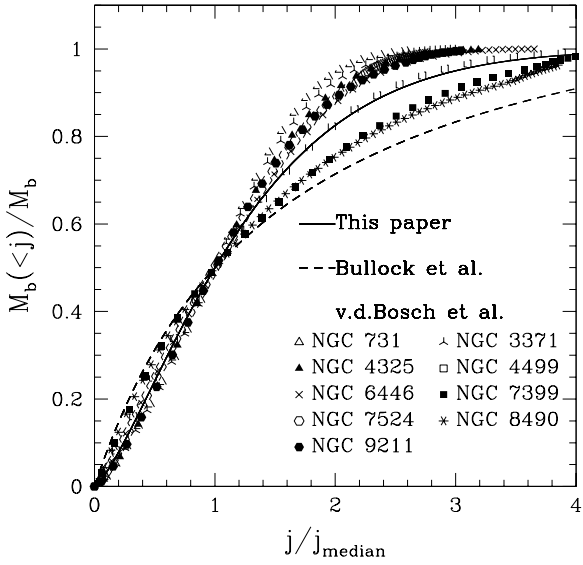}
\caption{Baryon mass profile according to our angular momentum
  prescription (solid line, cf eq. \protect{\ref{Mbj}} with $\gamma=1$
  and $n=4/3$), in comparison with the universal profile (dashed line)
  and the observed disk galaxies (symbols), where $j_{\rm median}$ is
  defined such that 50\% of the disk mass has the specific angular
  momentum $j<j_{\rm median}$.  The observed disk profiles are adopted
  from \protect{\citet{vdB:01}} with the stellar mass-to-light ratio
  $M/L_R = 1 \, M_{\sun}/L_{R,\sun}$.}
  \label{fig:expj}
\end{figure}

2) 
Here we also take another approach, assuming the disk mass
$M_b(r_j)$ can be expressed as a function $M_b(<j(r_i))$ of the
initial specific angular momentum $j(r_i)$, where $j(r_i)$ is given by
equation (\ref{j}).  The advantage is that the r.h.s. of equation (\ref{rj})
can now be expressed as an explicit function of $r_i$, and hence $r_j$
is computed directly without iteration as in the previous approach.

Since the $z$-component of the specific angular momentum originated
from tidal torquing of neighbouring halos, hence it was distributed in
the same way in baryons and dark matter, we could adopt the so-called
universal angular momentum profile, $M_b(<j)=f_b M_{\rm vir} \, j/(0.8
j+0.2 j_{\rm max})$, seen in dissipationless N-body simulations of
$\Lambda$CDM models (\citealt{Bullock:01b}), where $j_{\rm max}$ is
the maximum specific angular momentum determined by the spin parameter
$\lambda$.  Unfortunately, if baryons with such a distribution
collapse to a disk conserving angular momentum, then the central
density of the disk would be too high and the scale length too low
compared to the observed disks \citep{Bullock:01b,vdB:01}.  The
angular momentum distribution of such a disk is also very different
from that of exponential disks, as shown in Fig. \ref{fig:expj}.  This
mismatch between the present day exponential disks and the simulated
initial baryon angular momentum distribution is an intriguing problem,
which might argue for the transfer of angular momentum by bars or
spiral waves in the disk or unvirialised lumps in the halo.  We will
not digress into this issue since the goal of our analytical and
N-body modeling is to explore the maximum effect of blowing out the
exponential disks.

We have decided to make an alternative choice of the functional form
of the initial cumulative distribution of the specific angular
momentum of baryons,
\begin{equation}\label{Mbj}
{M_b(<j) \over f_b M_{\rm vir}}  =1-\exp\left({-{j^n \over j_0^n}}\right),\qquad
n \equiv {6-2\gamma \over 4-\gamma},
\end{equation}
where $1.33 \ge n \ge 1.2$ for a general cusp slope $1 \le \gamma \le
1.5$ of the initial halo, and $j_0$ is a characteristic specific
angular momentum, related to the median angular momentum by $j_{\rm
median} \equiv j_0 (\ln 2)^{1/n}$.  This parametrization can describe
the angular momentum profile of observed disks fairly well (cf
Fig. \ref{fig:expj}), albeit with significant scatter.  It also
creates an approximately exponential disk after adiabatic contraction.
This can be understood roughly since $j = r_j v_j \propto r_j$ in a
model with a nearly flat rotation curve, so that equation (\ref{Mbj})
corresponds to an almost exponential distribution in radius $r_j$.
Also, our angular momentum profile reproduces the asymptotic profile
$M_b \propto j^{4/3} \propto r_j^2$ near the center, as expected for
the exponential disk embedded in a Hernquist halo with $\gamma=1$ and
$n=4/3$.  The characteristic specific angular momentum $j_0$ is
determined by the integrated total angular momentum $J$ of the initial
halo,
\begin{equation}
\int_0^\infty \!\!\!\!\!\! dx_j x_j 
{d \left[1-\exp\left(-x_j^n\right)\right]\over dx_j} 
={J \over M_{\rm vir} j_0 } 
={\lambda G M_{\rm vir}^{3/2}|E|^{-1/2} \over j_0},
\end{equation}
where the l.h.s. can be reduced to the gamma function, $\Gamma(1+1/n)$.
Since the baryons have initially the same distribution as dark matter,
the total energy of the system is
\begin{equation}
|E| = (1-f_b)^{-2} \int_0^{\infty} dr {dM_{dm} \over dr} {GM_{dm} \over 2r} = 
      {GM_{\rm vir}^2 \over 4 (5-2\gamma) r_s},
\end{equation}
and we can express $j_0$ in terms of the $\lambda$ parameter, 
\begin{equation}
j_0 = {\lambda \sqrt{G M_{\rm vir} r_s} \over \xi},
      \qquad \xi \equiv \Gamma(1+1/n)/\sqrt{4(5-2\gamma)}.
\end{equation}

At small radii the initial mass distribution of dark matter is
\begin{equation}
M_{dm} \rightarrow (1-f_b) M_{\rm vir} \left({r_i \over r_s}\right)^{3-\gamma}, 
  \qquad r_i \rightarrow 0,
\end{equation}
and the initial specific angular momentum
\begin{equation}
j \propto \sqrt{M_{dm}r_i} \propto r_i^{4-\gamma \over 2}
  \propto M_{dm}^{4-\gamma \over 6-2\gamma}.
\end{equation}
So with our choice of $n$, the asymptotic angular momentum distribution
of the baryons scales the same way as that of the dark matter:
\begin{equation}
M_{dm} \propto M_b \propto j^n, \qquad j \rightarrow 0.
\end{equation}
It follows then that the mass profiles of the baryons and dark matter
are similar before and after the contraction,
\begin{equation}
M_b \propto M_{dm} \propto r_i^{3-\gamma} \propto j^n \propto r_j^{3-\gamma} ,
\end{equation}
where 
\begin{equation}
r_j \propto j^{2 \over 4-\gamma} \propto r_i
\end{equation}
is the post-contraction radius.  The contraction factor is finite at
small radii, and the cusp slope of the dark matter does not change.  To
summarize, at small radii after the contraction we have
\begin{eqnarray}\label{eq:smallmdm}
M_{dm} &\rightarrow & (1-f_b) M_{\rm vir} \left({C r_j \over r_s}\right)^{3-\gamma}, \\
M_b    &\rightarrow & \xi^{n} \lambda^{-n} f_b M_{\rm vir}\left({C r_j \over r_s}\right)^{3-\gamma}, \\
r_j    &\rightarrow & C^{-1} r_i,
  \label{eq:rj}
\end{eqnarray}
where
\begin{equation}
C \equiv 1+(\xi^{n} \lambda^{-n}-1) f_b.
  \label{eq:C}
\end{equation}
At a given radius $r_j$, the enclosed dark mass and density are enhanced
by a factor $C^{3-\gamma}$.  For an initial Hernquist halo, $\gamma=1$,
$n=4/3$, and $\xi=0.265$.  For the disk parameters $f_b=0.13$,
$\lambda=0.05-0.01$, the inner dark matter density is enhanced by a
factor $C^2=4-160$.

\begin{figure}
\epsfxsize=8.4cm
\epsfbox{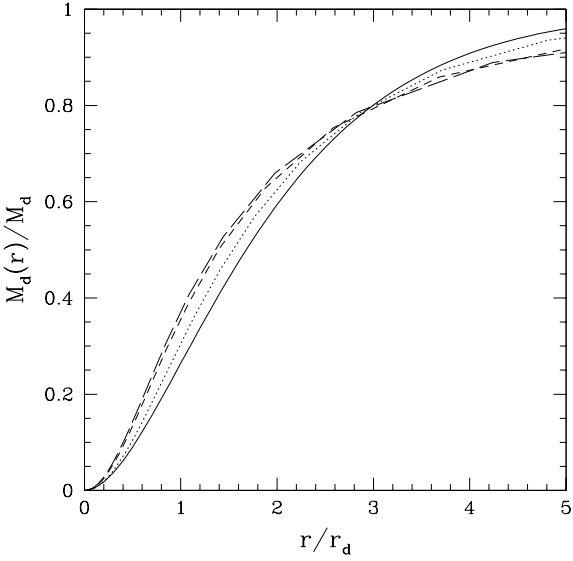}
\caption{Baryon mass profile according to our angular momentum
prescription, in comparison with the exponential disk profile (solid
line; cf. eq.~\protect{\ref{exp}}).  The radii are normalized to the
scale length $r_d$ computed numerically for three cases: $\lambda =
0.05$ (dots), $\lambda = 0.01$ (short dashes), $\lambda = 0.0025$ (long
dashes).  \label{fig:diskprofile} }
\end{figure}

It is interesting to note that the model with $\gamma=1$ produces an
asymptotic mass distribution at small radii, $M_b \propto r_j^2$,
similar to that of an exponential disk (cf eq. \ref{exp}).  Figure
\ref{fig:diskprofile} shows the computed mass profiles of baryons for a
family of models with different $\lambda_b$.  Our analytical models
reproduce exponential profiles at 10\% level within $0.5-5$ scale
lengths.  The difference in the disk potential is even less and
therefore the adiabatic contraction of dark matter is very similar.  For
all three models, the dark matter profiles computed with the two methods
are essentially indistinguishable between $10^{-3}$ and $10 \, r_s$.

\subsection{Analytical models of slow wind and fast wind}
  \label{sec:anamodel}

The expulsion of gas can be worked out analytically in two limiting
cases: the adiabatic and instantaneous regimes.  We shall refer to these
as the slow wind and the fast wind, respectively.  The outflow can play
a significant role in massive galaxies as well as in dwarf galaxies.
Our analytical description is similar to that used in estimating the
maximum wind-induced expansion of the dark halo of the Milky Way
(\citealt{Zhao:02}).

To gain more analytical insight, we first consider the slow wind.  In
this case, a circular orbit of radius $r_j$ expands to a circular orbit
of radius $r_f$ while conserving the specific angular momentum
$\sqrt{GM(r)r}$.  So the expansion factor is inversely proportional to
the combined enclosed mass of baryons and dark matter, i.e.,
\begin{equation}
{\left. {r_f \over r_j}\right|}_{\rm slow} =  
  {M_{dm}(r_j) + M_b(r_j) \over M_{dm}(r_j) + M_b(r_j) - M_\wind(r_j)}
\end{equation}
where $M_b(r_j)$ and $M_{dm}(r_j)$ are the baryon and dark matter masses
within radius $r_j$, and $M_\wind(r_j)$ is the amount of baryons
``gone with the wind''.  Alternatively, we can write
\begin{equation}\label{eq:rfslow}
{\left. r_f\right|}_{\rm slow} ={r_j \over 1-f_\wind(r_j)},
\end{equation}
where
\begin{equation}\label{eq:fwind}
f_\wind(r_j) \equiv {M_\wind(r_j) \over M_{dm}(r_j) + M_b(r_j)},
\end{equation}
measures the blown-away mass as a fraction of the dynamical mass.  The
expansion factor $r_f/r_j$ is larger in the inner region where the
baryons are concentrated and is minimal at the virial radius, where we
have only a modest change, since
\begin{equation}
f_\wind(r_{\rm vir}) = \delta f_b \, \la \, 0.1,
\end{equation}
where $0 \le \delta \le 1$ is the total fraction of baryons blown away.

Although we do not know of any close form for the effect of fast wind, 
we could modify equation (\ref{eq:rfslow}) for slow wind
to mimic the effects of wind of arbitrary strength.
We suggest the following simple formula
\begin{equation}
r_f = { {\left. r_f\right|}_{\rm slow} \over
\left[1-f_\wind(r_j)\right]^{k \over 4} (1-2^k\delta f_b)^{k \over 4} },
  \qquad  0 \le k \le 1,
  \label{eq:rf}
\end{equation}
where $k$ is a tunable parameter to model the rapidity of the wind.
For example, slow wind is a special case with $k=0$ (cf
eq. \ref{eq:rfslow}).  The instantaneous wind is modeled with $k=1$,
following the rule of thumb that a system becomes unbound if it loses
half of the mass instantaneously.  To verify this, we substitute $k=1$
and $f_\wind(r_{\rm vir}) \rightarrow 1/2$ in equation (\ref{eq:rf})
and find that the r.h.s. factor $1 - 2^k \delta f_b = 1 - 2
f_\wind(r_{\rm vir}) \rightarrow 0$, hence $r_f \rightarrow \infty$.
For the winds of moderate speed with $k<1$ and moderate mass-loss with
$f_\wind(r_{\rm vir}) < 2^{-k}$, the system remains bound and expands
by a certain factor at large radii.  At small radii, where the baryons
always dominate with $f_\wind(r_j) \rightarrow 1$, the particles may
not necessarily escape to infinity because they slow down when they
cross the orbits of outer particles.  As we will show in
\S\ref{sec:nummodel}, this empirical parametrization turns out to
approximate well the results of numerical simulations.

To work out the properties of the model at very small radii, we
substitute equations (\ref{eq:smallmdm}-\ref{eq:C}) in equation
(\ref{eq:fwind}) and obtain
\begin{equation}
f_\wind \rightarrow \xi^n \lambda^{-n} \delta \, C^{-1} f_b.
  \label{eq:smallr}
\end{equation}
Substituting equations (\ref{eq:rj},\ref{eq:smallr}) in equation
(\ref{eq:rf}), we find that in the center the final post-wind radius
$r_f$ is related to the initial pre-contraction radius $r_i$ by
\begin{equation}  \label{eq:Fdef}
F \equiv {r_f \over r_i} = 
  {\left[1+(\xi^n \lambda^{-n}-1)f_b\right]^{k \over 4} \over 
  \left[1+(\xi^n \lambda^{-n}(1-\delta)-1)f_b\right]^{1+{k \over 4}}(1-2^k\delta f_b)^{k \over 4}}.
\end{equation}
The slope of the cusp remains the same.  By comparing the dark matter
mass within a given radius, we find that the density is reduced by a
factor $F^{3-\gamma}$.

\subsection{Maximum effect on the final halo density}
  \label{sec:finalmodel}

The strongest effect on the halo is achieved in the limit of complete removal
of the baryonic disk, i.e., $\delta \rightarrow 1$.  In this case we have
for all radii
\begin{equation}\label{eq:finalinit}
r_f = {r_i  \over \left(1-f_b\right)} 
  \left[{1+M_b/M_{dm} \over 1-2^{k}f_b}\right]^{k \over 4}, 
  \qquad \delta \rightarrow 1.
  \label{eq:rf1}
\end{equation}
Interestingly, 
if the disk forms slowly and is removed slowly as well (i.e., with $k=0$),
the halo expands slightly by a constant factor
\begin{equation}\label{eq:slowslow}
{r_f \over r_i} = \left(1-f_b\right)^{-1} = const,
\end{equation}
as expected from adiabatic invariance.  The result is the same if all
the baryons were removed from the initial halo distribution without
going through the phase of disk formation.

The effect is maximized when a very dense disk ($\lambda \ll 0.05$) is
removed instantaneously ($k=1$).  If the initial halo is a Hernquist
model (i.e., $\gamma=1$, $n={4 \over 3}$, and $\xi=0.265$), then close
to the center the halo will expand by a factor (cf eq. \ref{eq:Fdef})
\begin{equation}
F \sim {1 \over 1-f_b}
\left[{f_b \over (1-f_b)(1-2f_b)}\right]^{1 \over 4} 
\left({0.265 \over \lambda}\right)^{1 \over 3}
\sim \left({0.15 \over \lambda}\right)^{1 \over 3},
  \label{eq:F}
\end{equation}
relative to the original halo model.  Here we assume a universal
baryon fraction $f_b=0.13$ to make a massive disk.  To summarize, the
density of the innermost dark matter can drop at most by a factor
$F^2=2-6$ for $\lambda=0.05-0.01$.

\section{Spherical shell simulations}
  \label{sec:nummodel}

To confirm the analytical results, we compute the dynamics of dark
matter using the spherical shell method of \citet{GO:99}.  The
algorithm, originally due to L. Spitzer and collaborators, utilizes
spherical geometry to achieve high resolution in the center.  Dark
matter particles are distributed on concentric spherical shells at radii
$r_k$ with the mean energy $E_k$ and angular momentum $j_k$.  Shells
move in the radial direction with the velocity $v_{r,k}$ and can freely
cross each other.

We have extended the code of \citet{GO:99} by including variable shell
masses, $m_k$, and improving the corrections for shell crossing.  The
code solves the equations of motion using the first integral, $C_k$:
\begin{equation}
  v_{r,k}^2 + {j_k^2 \over r_k^2} - {2 G M(r_k) \over r_k} = C_k,
\end{equation}
where $M(r_k)$ is the mass enclosed within shell $k$, including a half
of its own mass (accounting for the shell self-gravity):
\begin{equation}
  M(r_k) = \sum_{j=1}^{k-1} m_j + {m_k \over 2},
\end{equation}
if the shells are ordered in radius.  The integrals $C_k$ determine
the turn-around points where the direction of the shell's velocity is
reversed.  At each time step, the code finds the locations of all
shell crossings and adjusts the integrals $C_k$ according to the
changed enclosed mass.  Because of such corrections, two-body
interactions are effectively reduced and there is no need for force
softening.  As a result, the central region of the system can be
probed with very high accuracy.

In order to investigate the effect of angular momentum loss, we have run
three models: Model A with the mean value of $\lambda = 0.05$, and
Models B and C where $\lambda$ deviates from its mean by $3 \,
\sigma_\lambda$ ($\lambda = 0.01$, Model B) and $6 \, \sigma_\lambda$
($\lambda = 0.0025$, Model C).  

We use the units $G = 1$, $M_{\rm vir} = 1$, $r_s = 1$, and therefore
the results can be scaled to any halo parameters.  Each model is run
with $N = 10^4$, $4\times 10^4$, and $10^5$ shells to check numerical
convergence.  After the disk is removed, the models are run for 40
half-mass dynamical times ($t_{\rm dyn} = 8$ in code units) to ensure
that a new dynamical equilibrium is achieved.

\begin{figure}
\epsfxsize=8.4cm
\epsfbox{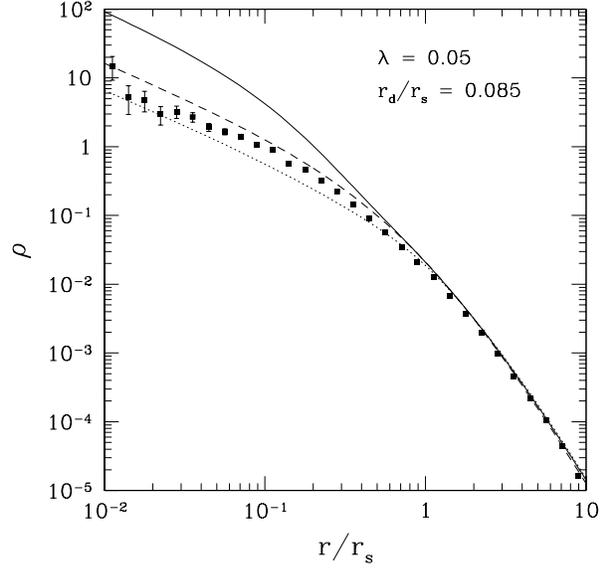}
\caption{Dark matter density profiles for Model A ($\lambda=0.05$,
  $f_b=0.13$, $\delta=1$) before and after the gas removal.
  Dashed line is the initial Hernquist model, solid line is
  the halo profile after the adiabatic contraction.  Symbols
  show a new equilibrium profile after the removal of the
  baryonic disk, computed with $10^5$ spherical shells.
  Errorbars are the sampling uncertainty.  Dotted line is
 the analytical prediction, eq. (\protect\ref{eq:rf1}).}
\label{fig:ModelA}
\end{figure}

\begin{figure}
\epsfxsize=8.4cm
\epsfbox{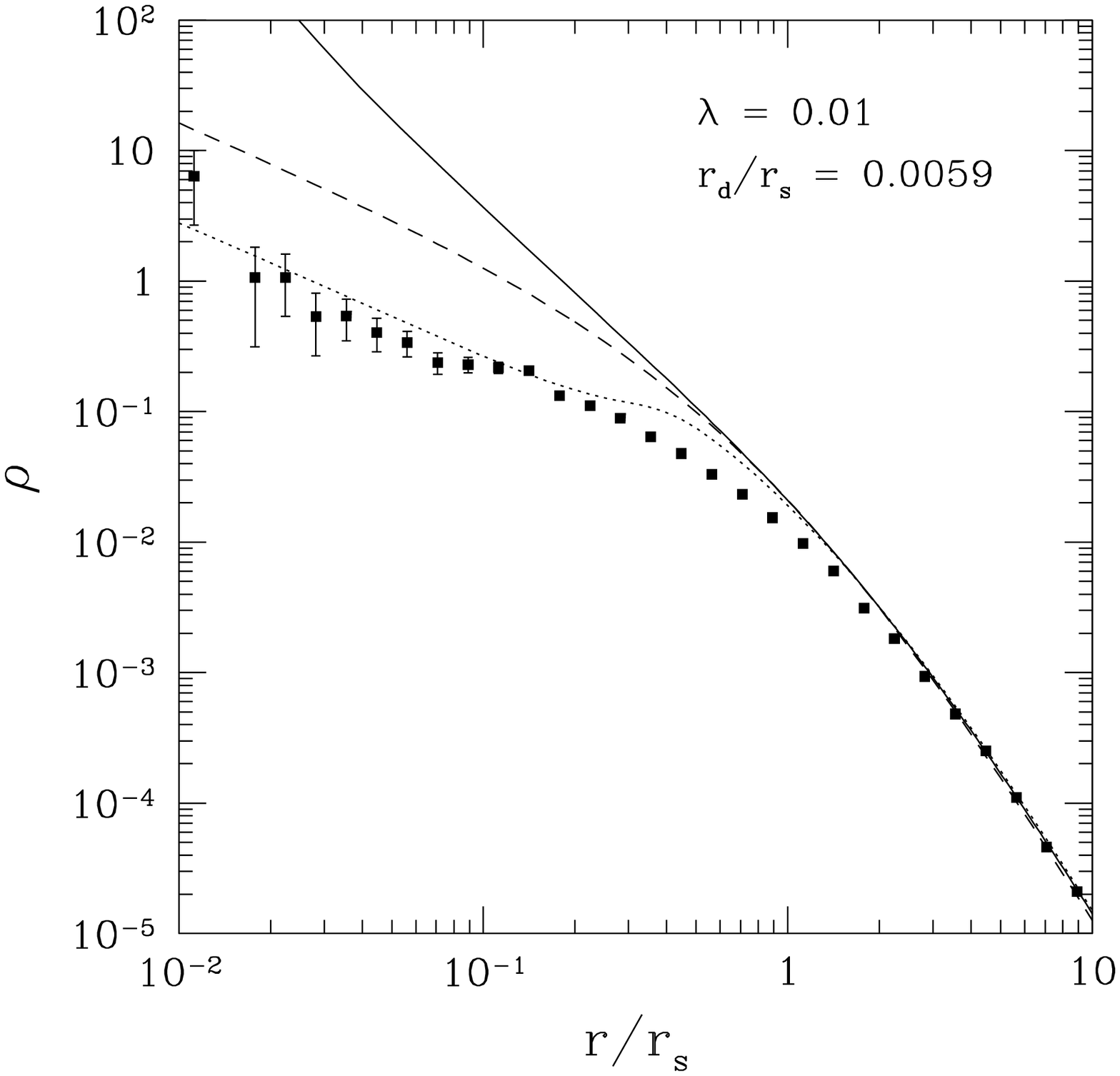}
\caption{Density profiles for Model B ($\lambda=0.01$).
  Lines are as in Figure \protect\ref{fig:ModelA}.}
\label{fig:ModelB}
\end{figure}

\begin{figure}
\epsfxsize=8.4cm
\epsfbox{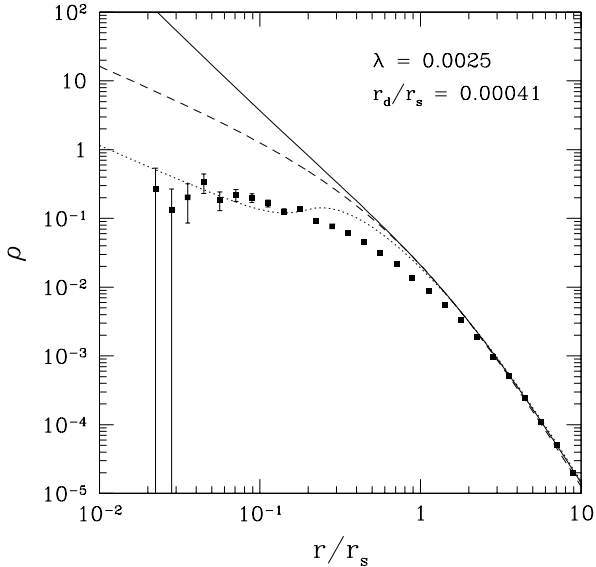}
\caption{Density profiles for Model C ($\lambda=0.0025$).
  Lines are as in Figure \protect\ref{fig:ModelA}.}
\label{fig:ModelC}
\end{figure}

Figure \ref{fig:ModelA} shows the initial halo profile and the new
equilibrium distribution after the instantaneous removal of the whole
disk in Model A.  The scale length of the disk, $r_d = 0.085\, r_s$,
is already a small fraction of the scale radius of the halo.  As
expected, there is no change in the profile outside the cusp, at $r >
r_s$.  But even inside the scale radius, the new profile is fairly
close to the initial model prior to adiabatic contraction.  The
central density is affected at most by a factor of 2, in agreement
with the analytical prediction (eq. \ref{eq:F}).

Figure \ref{fig:ModelB} shows the density profile for Model B.  The
effect is stronger but there is still no indication of a core forming
instead of the cusp.  The asymptotic solution increases towards the
center as $r^{-1}$ and describes the numerical result reasonably well.

Finally, Model C may have achieved a central core (Figure
\ref{fig:ModelC}).  The inner profile can be fitted with an
approximate isothermal sphere, $\rho \propto (r^2 + r_c^2)^{-1}$,
where $r_c \approx 0.2\, r_s$.  This core radius is significantly
larger than the tiny initial scale-length of the disk, $r_d =
0.00041\, r_s$.  However, it seems that the energy input has already
saturated in Model C (the change in the binding energy is similar to
Model B) and further contraction of the disk would not lead to a
larger core.

After the disk is removed, the dark matter particles with a large
enough kinetic energy become unbound.  At the end of the simulation,
the halo in Model A loses 6.5\% of its mass, while halos B and C lose
14\% and 16\%, respectively.  (About half of those particles become
unbound during the subsequent evolution of the halo potential towards
a new equilibrium.)  However, as the disk is made progressively more
compact, the heating caused by the disk removal is shared by a small
number of particles in the center.  They reach very high velocities
and leave the system, whereas most of the particles remain unaffected.
In the limit of an infinitely small disk, the final equilibrium
profile would still be close to our extreme Model C.

\subsection{Artificial numerical effects}

As any N-body simulation, our models are affected to some extent by
the artificial numerical relaxation.  This relaxation arises from the
small-scale particle encounters and from the Poisson potential
fluctuations due to discreteness \citep{HB:90,W:93}.  Even though the
shell-crossing corrections in our code remove the small-scale errors,
the large-scale fluctuation modes are unavoidable.

The relaxation time over which the particle energy changes by the
order itself can be estimated at the half-mass radius of the system,
$R_h$, from \citep{SH:71}:
\begin{equation}
  t_r \approx 0.14 \, {N \over \ln{\Lambda}} 
              \left({R_h^3 \over GM}\right)^{1/2},
\end{equation}
where $\ln{\Lambda} \sim \ln{N}$ is the usual Coulomb logarithm.  In
the code units, $G=1$, $M=1$, and with $R_h \approx r_s = 1$, we have
\begin{equation}
  t_r \sim 0.1 \, {N \over \ln{N}}.
    \label{eq:tr}
\end{equation}

This expression can be compared with the relaxation rate (inverse of
the relaxation time) derived by \citet{W:93}.  His Fig. 3 expresses
$t_r$ as a function of the system size divided by the Jeans length.
For a virialised Hernquist model, this ratio can be $0.6 - 0.8$ and
reading of the plot gives $t_r \sim 25 - 50$.  It agrees with equation
(\ref{eq:tr}) for the particle number $N = 4000$ used by \citet{W:93}.

Thus, the large-scale potential fluctuations can only be reduced by
increasing the number of particles in the simulation.  We estimate
that our results with $N = 10^5$ shells are unaffected by the
relaxation at the scale radius $r_s$ for $t \ll t_r \approx 800
\approx 100 \, t_{\rm dyn}$.  The rate of relaxation does not increase
much in the inner parts, because it is dominated by the global modes
and not by local encounters.  

Another numerical effect, which is more important for the inner shells
than the outer shells, is the kinematic error due to a finite time
step.  We chose $\Delta t = t_{\rm dyn}/N$, so that each shell crosses
on average one other shell.  Since the density at $r < r_s$ goes
approximately as $\rho(r) \propto r^{-1}$, the orbital period of a
shell $t_{\rm orb}(r) \approx t_{\rm dyn} \, (r/r_s)^{1/2}$.  This
shows that if 10 time steps are necessary to properly resolve an
orbit, our results are valid down to a tiny radius $r \sim (10/N)^2 \,
r_s \sim 10^{-8} \, r_s$.

\subsection{Comparison with previous work}

\citet{NEF:96} studied the effect of instantaneous disk removal using
a three-dimensional N-body code.  They also used a Hernquist model for
the initial dark matter profile and imposed an external potential of
the exponential disk.  Instead of analytically calculating the
adiabatic contraction, they let the dark matter halo to adjust
dynamically to the disk potential.  Then, they removed the disk and
studied the subsequent expansion of the halo.  Navarro et al. found
that the core did develop in the inner regions of all of their models.
They fit a non-singular isothermal sphere to the inner 25\% of the
mass and found the core radii satisfying the following relation: $r_c
= 0.11 \, (M_d/r_d)^{1/2}\, r_s$, where $M_d$ is the mass of the disk
in units of the halo mass (or $f_b/(1-f_b)$, in our notation).

The expansion effect that Navarro et al. find is much stronger than
what our simulations or analytical modeling show.  For example, their
fit for the core radius would give for our Model C, $r_c \approx 2.1$,
about ten times the value we find.  In Models A and B the core radii
should also have been easily detectable, according to the fit.

We identify two possible reasons for the disagreement, force
resolution and numerical relaxation effects.  Firstly, a force
softening at 0.03 $r_s$, used by \citet{NEF:96}, could prevent an
accurate calculation of the dynamics of inner particles, especially in
cases of very small disks.  This would naturally cause a development
of a core in the particle distribution, although the core radii found
by Navarro et al. are larger than the softening length.  Secondly,
numerical effects might be important, as their simulations use only $N
= 10^4$ particles.  The relaxation time, $t_r \sim 100$ time units, is
4 to 6 times shorter than the duration of their experiments.  Even
though Navarro et al. demonstrate that the numerical profile before
the disk removal is stable, it may be more susceptible to the
relaxation afterwards, when a smaller number of particles is left in
the center.  This may also lead to a simple statistical undersampling
of the density distribution, posing as a 'core'.

It is likely that the difference in the results is not related to the
geometry, 1D vs 3D simulations.  In our one-dimensional models, all
shells conserve their initial angular momentum and do not experience
any non-radial perturbations.  But since the dark matter halo is not
centrifugally supported, precise conservation of angular momentum
should not seriously affect the dynamics.  Also, the remaining stellar
distribution in dwarf spheroidal galaxies is observed to be roughly
spherical, so we can expect the dark matter halo to be spherical as
well, lending support to our 1D models.  Finally, we have checked that
our different implementations of the adiabatic contraction gave almost
identical initial conditions and therefore cannot be responsible for
the disagreement.

Note, that our simulations achieve much higher resolution, with ten
times more particles and no force softening in the center.  But the
statistical error-bars in Figure \ref{fig:ModelC}, inversely
proportional to the square root of the number of particles in the bin,
show that even our models cannot probe regions smaller than $0.01 -
0.03 \, r_s$.  At larger radii, except for the extreme model C, we do
not find any significant flattening of the dark matter distribution.

Any realistic modeling of stellar feedback would have an even weaker
effect.  \citet{LCS:00}, using a more complex tree-SPH simulation of
star formation in a dwarf galaxy, find that the central cusp remains
intact.

\subsection{Application to DDO 154}

\begin{figure}
\epsfxsize=8.4cm
\epsfbox{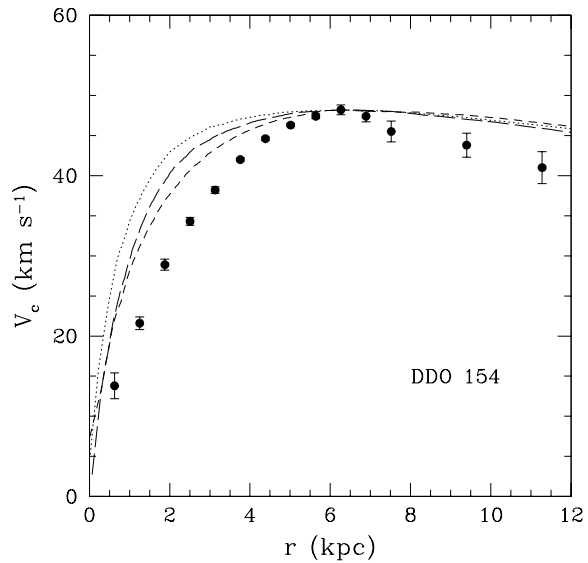}
\caption{Rotation curve of DDO 154 from a high-resolution HI data
(from \protect\citealt{CP:98}) compared with the final circular
velocities of Models A (dots), B (short dashes), and C (long dashes).
The simulated rotation curves are normalized to pass through the
maximum of the observed one, with no formal fitting attempted.}
\label{fig:ddo154}
\end{figure}

Our results demonstrate that stellar feedback is insufficient to
reduce the central dark matter density.  We illustrate this point for
a prototypical low surface brightness galaxy DDO 154.  Figure
\ref{fig:ddo154} shows the observed rotation curve of DDO 154, which
peaks at 48 km s$^{-1}$.  As this is perhaps the most robustly
determined quantity, we normalize the simulation results to have the
same value and location of the circular velocity.  The simulated
curves correspond to the final equilibrium density profiles of Models
A, B, and C.  They are clearly more concentrated than allowed by the
data.

In addition, the scale radii of the initial halo in Figure
\ref{fig:ddo154} increase as the effect of feedback gets weaker
(corresponding to the larger disk).  For Model C $r_s = 2.9$ kpc, but
for the less perturbed models B and A it is $r_s = 3.6$ and 5.6 kpc,
respectively.  The virial halo mass is about the same, around $1.3
\times 10^{10}\, M_{\sun}$.  This mass is lower than the best-fitting
value ($\sim 3\times 10^{10}\, M_{\sun}$) from \cite{vdB:00}, and yet
the resulting rotation curve is well above observations.

Note, that we do not attempt any accurate fitting of the rotation
curve.  The beam-smearing effects are likely to be unimportant (beam
size is less than 1 kpc; \citealt{vdB:00}), and therefore the small
error-bars in the inner parts place very strong constraints on the
allowed dark matter profiles.  \citet{GS:99} used a set of N-body
simulations to address this issue.  They selected three halos with $M
\sim 10^{10}\, M_{\sun}$ at $z=0$ from a random realization of a $4 \,
h^{-1}$ Mpc box with the SCDM power spectrum.  Varying the amount of
gas and the concentration of the dark matter halo, \citet{GS:99} could
find a satisfactory fit to the inner rotation curve of DDO 154 only
(1) if the present disk contained 3 times more gas than is observed
and (2) if the halo had a low concentration, $r_{\rm vir}/r_s = 4$.
However, none of their models could fit simultaneously the inner and
outer parts of the rotation curve.

\section{Summary}
  \label{sec:summary}

We have explored the effect of maximum feedback on the central density
of dark matter halos of the gas-rich dwarfs, combining analytical
limits with numerical simulations.  The expansion of dark matter after
the removal of the entire disk is controlled by the mass and
compactness of the disk.  For a wide range of the baryon angular
momenta, we find the effect to be modest, at most a factor of 2 to 6
reduction in the central halo density.  This is hardly enough to bring
the models into agreement with the observed solid-body rotation
curves, as we demonstrate for the case of DDO 154.

We conclude that the slowly rising rotation curves are likely to be a
genuine problem of CDM models.  It might be necessary to consider
other possible solutions, such as the effect of merging of massive
black holes or an unusual property of dark matter particles.

\bigskip\noindent 
{\bf Acknowledgements} \\
We would like to thank Martin Rees for discussions, and Frank van den
Bosch for comments and for making available the data used for
generating Fig. \ref{fig:expj}.

\end{document}